# Pengaruh Model Pembelajaran Generatif Berbantuan Media *Mind Mapping* terhadap Kompetensi Fisika Siswa Kelas XI MIA SMAN 1 Painan

**Dila Ayu Lestra[1], Akmam[2], Yenni Darvina[3], Fuja Novitra[4]**

[1,2,3,4] Fisika, Universitas Negeri Padang

e-mail: dilaayulestra@gmail.com

**Abstrak**

Siswa kurang aktif dan kritis pada proses pembelajaran. Siswa sekedar menyimak apa yang guru jelaskan tanpa memahami isinya, dan melupakannya pada pembelajaran berikutnya sehingga mengurangi efektivitas pembelajaran. Solusi yang mungkin dilakukan yakni melalui penerapan model ajar generatif dengan bantuan media *mind mapping.* Tujuan dilaksanakannya penelitian yakni mengetahui pengaruh dari model ajar generatif yang dibantu dengan media *mind mapping* ditinjau dari kompetensi fisika siswa Kelas kelas XI MIA SMAN 1 Painan. Hipotesis penelitian ini yakni ditemukan pengaruh dari penerapan model ajar generatif yang dibantu dengan media *mind mapping* ditinjau dari hasil belajar materi dinamika gerak. Penelitian ini berjenis eksperimen semu. Populasi untuk penelitian yakni keseluruhan siswa kelas XI FASE SMAN 1 Painan tahun akadmik 2023/2024. Teknik *purposive sampling* dipakai untuk pengambilan sampel. Sampel penelitiannya yakni Kelas XI FASE 1 dan Kelas XI FASE 2. Instrumen yang digunakan berupa lembar berupa pertanyaan objektif. Uji-t digunakan untuk analisis data, dan tingkat signifikansinya adalah 0,05. Setelah dilakukan penelitian, kedua kelas sampel tersebut diberi perlakuan yang berbeda. Penelitian memperlihatkan ditemukannya perbedaan yang signifikan dari kelas yang menerapkan model ajar generatif dibantu dengan media *mind mapping* terhadap kelas yang menerapkan model ajar yang biasa guru pakai melalui bantuan media *mind mapping.* Nilai belajar siswa rata-rata untuk kelas yang memakai model ajar generatif diraih 79,20, dan nilai belajar siswa rata-rata untuk kelas yang memakai model ajar yang biasa dipakai pendidik diraih 76,15. Dampak dari model ajar generatif yang dibantu dengan media *mind mapping* terlihat pada hasil proses belajar siswa yang dianalisis dan diuji hipotesisnya. Berdasarkan uji hipotesis diperoleh $t_{tabel}$ adalah 1,09 dan $t_{hitung}$ adalah 2,00. Syarat $H_0$ ditolak adalah jika $t_{tabel} < t_{hitung}$. Nilai $t_{hitung}$ berada dalam penolakan $H_0$, sehingga $H_1$ diterima. Karena semua variabel dikontrol, kecuali model proses belajar sehingga bisa diraih Kesimpulan yakni penerapan model ajar generatif yang dibantu dengan media *mind mapping* pada materi Dinamika Gerak menyumbangkan pengaruh positif untuk hasil belajar peserta didik.

**Kata kunci:** *Model Pembelajaran Generatif, Media Mind Mapping, Kompetensi Fisika*

**Abstract**

Students are less active and critical in the process of learning. Students pasif shown from just listen to the teacher's give explanation without understanding the contents, and forget it in the next lesson, thus reducing the effectiveness of learning. A possible solution is the a generative learning model application supported by mind mapping media. The research purpose was to reveal the generative learning model impact supported by mind mapping media on the physics competence of students in Class XI MIA SMAN 1 Painan. The research hypothesis is that there is an impact from the generative learning model application suppoerted by mind mapping media on the learning outcomes of motion dynamics material. This research type included in quasi-experimental research. The research population was all students of class XI FASE SMAN 1 Painan in the 2023/2024 academic year. The technique of purposive sampling was used for sampling. The research samples were Class XI FASE 1 and Class XI FASE 2. The instrument used was a sheet in the shape of objective questions. The t-test was utilized for analysis of data, and the level of significance was 0.05. After the research was conducted, the two sample classes were given different treatments. The research shown that there was a significant difference from classes using generative learning models supported by mind mapping media and classes using learning models commonly used by educators with the support of mind mapping media. The average student learning score in classes using generative learning models was 79.20, and the average student learning score in classes using learning models commonly used by educators was 76.15. The impact of the generative learning model supported by media of mind mapping can shown in the learning outcomes of students who were analyzed and tested for their hypotheses. Based on the test of hypothesis, the t table was 1.09 and the t count was 2.00. The results for H0 to be rejected is if $t_{table} < t_{count}$. The t count value is within the rejection of $H_0$, so $H_l$ is accepted. Because all variables are controlled, except for the learning model, it can be said that the application of the generative learning model supported by mind mapping media on the Dynamics of Motion material has a positive effect on student learning outcomes.

**PENDAHULUAN**

Fisika diketahui sebagai ilmu yang mendalami seputar fenomena melalui serangkaian proses ilmiah berdasarkan bukti ilmiah dan mengimplementasikan hasilnya dalam bentuk produk ilmiah, dan terdiri dari tiga unsur utama: konsep, prinsip, dan teori yang akan dilakukan. Pelajaran Fisika adalah tentang benda-benda dan kejadian-kejadian di alam, dan sangar erat terhadap kehidupan keseharian dari manusia. Pendidikan fisika mempunyai tujuan yang lebih spesifik, yaitu membekali siswa dengan pengetahuan dan pemahaman untuk maju ke jenjang pendidikan yang lebih tinggi. Mempelajari fisika berarti membuat rumusan permasalahan, melakukan pengembangan dan pengujian hipotesis lewat uji coba, melakukan pengumpulan, pengolahan, dan penafsiran data, penyajian hasil uji coba lewat lisan atau tulisan, dan memakai konsep dan prinsip dari fisika. Tujuannya adalah untuk memperoleh pengalaman pada pemberian penjelasan bermacam-macam kejadian di alam.

Proses belajar Fisika bisa menunjang peningkatan keterampilan penalaran, pemikiran yang kritis dan kreatif melalui penggunaan konsep dan prinsip fisika dalam memberikan bermacam penjelasan kejadian di alam dan menuntaskan permasalahan dengan cara kualitatif dan kuantitatif. Pengajaran fisika diharapkan bisa memajukan ilmuwan dalam cara yang membantu mereka memahami konsep dan menerapkannya dalam kehidupan nyata. Siswa fisika diharapkan mampu memahami berbagai jenis fenomena alam. Hal ini tercapai apabila siswa mempunyai kemampuan berpikir yang baik. Mengingat pentingnya pembelajaran fisika, hendaknya siswa mengembangkan kemampuannya dalam hal sikap, pengetahuan dan keterampilan serta menikmati pembelajaran fisika.

Kenyataannya di bidang ini prestasi siswa dalam mata pelajaran fisika masih dikategorikan rendah. Hal tersebut memperlihatkan dimana tujuan pengajaran fisika belum diraih secara memadai. Berdasarkan observasi di SMAN 1 Painan diperoleh data rata-rata nilai ujian tengah semester (UTS) Kelas XI SMAN 1 Painan.
Hal ini ditunjukkan pada Tabel 1.

| No | Kelas | Jumlah Peserta didik (orang) | Nilai Rata-rata Ujian Tengah Semester 1 |
|---|---|---|---|
| 1 | XI FASE 1 | 30 orang | 75,60 |
| 2 | XI FASE 2 | 27 orang | 70,52 |
| 3 | XI FASE 3 | 28 orang | 67,57 |
| 4 | XI FASE 4 | 28 orang | 59,57 |
| 5 | XI FASE 5 | 30 orang | 62,13 |
| | Rata-rata | | 67,08 |

(Sumber: Pendidik Fisika SMAN 1 Painan)

Pada Tabel 1. Bisa diperhatikan rata-rata nilai UTS 1 pada pelajaran Fisika peserta didik dari kelas XI SMAN 1 Painan kebanyakan belum meraih angka Kriteria Ketuntasan Minimal (KKM) yang diatur sekolah yakni 78. Berdasarkan wawancara dengan pendidik Fisika di SMAN 1 Painan penyebab tidak optimalnya hasil pembelajaran Fisika adalah 1) persepsi awal peserta didik yang mempunyai anggapan pelajaran Fisika sukar dipahami, 2) pemahaman matematika dasar peserta didik lemah, 3) peserta didik kurang kritis dan hanya sedikit yang ikut serta dengan aktif pada proses belajar, ditandai dengan sedikitnya peserta didik yang bertanya pada pengajaran yang dilaksanakan, 4) peserta didik cenderung diam dan belum mampu menjawab pertanyaan ketika pendidik bertanya, 5) Siswa sulit mengingat apa yang telah dipelajari sebelumnya. Siswa berpartisipasi aktif pada proses belajar dan membentuk sebuah makna dari informasi di sekelilingnya, membangun pengetahuan dan pengalaman awal melalui penggunaan model ajar generatif dan media ajar berupa *mind mapping*.

　　　　Tahapan-tahapan model pembelajaran Generatif sangat cocok untuk mengatasi tidak optimalnya hasil pembelajaran Fisika. Persepsi awal siswa yang menganggap pelajaran fisika sulit dapat diatasi dengan tahap orientasi. Tahap orientasi bertujuan untuk membuat ketertsrikan perhatian dan motivasi dari peserta didik untuk topik yang akan dibahas secara eksplisit dan berfungsi sebagai aktivasi pengetahuan awal dalam upaya mencapai pembelajaran yang bermakna melalui proses kognitif dalam memori asosiatif. Peserta didik kurang kritis dan hanya sedikit yang ikut serta dengan aktif pada proses belajar, ditandai dengan sedikitnya peserta didik yang bertanya pada pengajaran yang dilakukan dapat diatasi dengan tahap konflik kognitif. Tahap konflik kognitif, yaitu tahap dimana peserta didik membangkitkan situasi konflik dalam struktur kognitif peserta didik. Pendidik memberikan pertanyaan yang bersifat menantang dan menggali untuk membantu peserta didik menemukan jawaban dugaan sementara. Peserta didik cenderung diam dan belum mampu menjawab pertanyaan ketika pendidik bertanya dapat diatasi dengan tahap pengungkapan. Tahap pengungkapan yaitu pendidik mengarahkan peserta didik untuk mengkonstruksi konsep sesuai dengan konsep-konsep ilmiah yang diajarkan dengan memberikan pertanyaan. Peserta didik mengalami kesulitan dalam mengingat kembali apa yang sudah dipelajari sebelumnya dapat diatasi dengan tahap konstruk. Tahap konstruk bertujuan agar peserta didik mencapai pemahaman konseptual dengan merumuskan kembali situasi, dan memecahkan masalah ambigu. Melalui tahapan yang ada pada model ajar generatif, siswa akan meraih kemampuan dan keterampilan dalam membentuk pengetahuan dengan mandiri.

　　　　Model belajar tipe generatif dikenal sebagia model yang memusatkan terhadap integrasi aktif pengetahuan dari pengetahuan awal dan isi pembelajaran melalui peran aktif siswa dalam pembelajaran. Pembelajaran ini berfokus pada pengintegrasian materi baru secara aktif ke dalam program siswa yang sudah ada (Maknun, 2015). Model pembelajaran terdiri dari unsur-unsur sebagai berikut: urutan tahapan proses belajar (sintaks), prinsip reaksi, sistem sosial, efektivitas/manajemen pendidikan, dan sistem yang sifatnya mendukung. Keempat komponen tersebut menjadi acuan praktis untuk guru dalam memakai sebuah model untuk proses belajar. Menerapkan model ajar tipe generatif masuk akal. Dampak tersebut mencakup dampak pembelajaran berupa hasil pembelajaran yang terukur dan dampak yang terkait, atau hasil pembelajaran (Irwandani, 2015).

　　　　Model pembelajaran generatif memiliki beberapa keunggulan yaitu, membuka ruang yang lebar supaya mengaitkan yang ada dipikirannya terhadap pengetahuan yang baru, serta melaksanakan diskusi dan menjumpai konsep baru pada bidang pengetahuan. Media ajar yang cocok dipakai melalui model ajar tipe generatif adalah media pemetaan pikiran. Media mind map dapat memperluas pengetahuan siswa. Media ajar berupa *mind map* merupakan metode belajar yang menampilkan informasi secara grafis. Media pembelajaran mind map dapat dipetakan melalui penggunaan garis percabangan, gambar, atau kata kunci yang berkaitan dengan konsep dan ide kunci.

　　　　*Mind mapping* dapat membantu siswa merencanakan, mengkomunikasikan, dan mengingat berbagai hal dalam banyak cara, membantu mereka memecahkan masalah dengan lebih kreatif, memusatkan perhatian, dan mengatur berbagai hal dengan lebih cepat dan efisien. Model ajargeneratif dan media ajar berupa *mind mapping* diterapkan pada mata pelajaran mekanika kinetik, karena TP 11.4 menganalisis hubungan gaya, massa, dan percepatan dalam konsep mekanika kinetik dan aplikasinya pada menyelesaikan permasalahan kehidupan keseharian disebabkan permasalahan yang ditemui pada bidang tersebut maka peneliti tertarik untuk melaksanakan penelitian yang berjudul "Pengaruh Model Pembelajaran Generatif Berbantuan *Mind Mapping* Terhadap Kompetensi Fisika Siswa Kelas Kelas XI MIA SMAN 1 Painan".

**METODE**

Jenis dari penelitian yang dipakai masuk dalam jenis quasi eksperimen. Walaupun desain ini mempunyai grup kontrol, tetapi tidak diberikan pengendalian secara penuh pada variabel eksternal yang memberikan pengaruh untuk proses penelitian [6]. Pendapat ini didukung oleh Yusuf [10] yang mengungkapkan desain ini tidak sekedar mendapat pengaruh oleh perlakuan saja tetapi juga oleh variabel lain.

Penelitian ini menerapkan metode eksperimen semu melalui penggunaan desain *post-test only*. Penelitian ini mencakup dua grup yakni eksperimen dan grup kontrol. Kedua grup ini akan meraih perlakuan berbeda. Grup kontrol akan meraih perlakuan dalam bentuk proses belajar memakai model ajar yang biasa diterapkan pendidik dengan bantuan media *mind mapping*, namun grup eksperimen akan meraih perlakuan dalam bentuk proses belajar melalui penggunaan media *mind mapping*. Model ajar tipe generatif yang dibantu dengan media *mind mapping*. Sesudah diterapkan perlakuan melalui model ajar dan media yang tidak sama, kedua grup akan melaksanakan tes akhir (*post-test*) guna meraih informasi hasil belajar siswa pada untuk tiap kelompoknya. Desain penelitian ini ditampilkan pada Tabel 2.

**Tabel 2. Desain *Post-test Only***

| Kelompok | Treatment | Post-test |
|---|---|---|
| Eksperimen | X | $T_2$ |
| Kontrol | - | $T_2$ |

(Sumber: Payadnya et al , 2018)

Keterangan:
X = Perlakuan yang diterapkan untuk grup eksperimen
- = Perlakuan yang diterapkan untuk grup kontrol
T1 = *Posttes* (tes akhir) pada grup eksperimen
$T_2$ = Posttes (tes akhir) pada grup kontrol

Populasi untuk penelitian ini yakni keseluruha siswa dari kelas XI FASE SMAN 1 Painan yang masuk daftar pada semester I tahun ajaran 2023/2024. Teknik pengambilan sampel menggunakan sampel yang ditargetkan. Targeted sampling adalah teknik untuk mengambil keputusan sampel dengan mempertimbangkan [7]. Dengan menggunakan sampel kelas yang gurunya sama dan rata-rata kelasnya sama, penulis bisa membentuk dua grup yakni eksperimen dari Kelas XI FASE 1 dan grup kontrol dari Kelas XI FASE 2.

**HASIL DAN PEMBAHASAN**
**Hasil**

Pada penelitian ini data hasil dari proses belajar Fisika Kelas XI SMAN 1 Painan diperoleh dari hasil *post-test* dua kelas sampel. Siswa mengerjakan tes dengan menjawab 25 soal pilihan ganda. Data uji pengetahuan siswa pada grup eksperimen dan kontrol ditampilkan pada Tabel 3.

| Kelas | N | Nilai Tertinggi | Nilai terendah | $\bar{X}$ | S | $S^2$ |
|---|---|---|---|---|---|---|
| Eksperimen | 30 | 96 | 60 | 79,20 | 10,63 | 112,99 |
| Kontrol | 27 | 92 | 56 | 76,15 | 9,44 | 89,21 |

Dari data yang diraih pada Tabel 3 memperlihatkan rata-rata nilai pengetahuan siswa pada untuk eksperimen lebih unggul dibanding dengan siswa pada grup kontrol. Hasil dari proses belajar siswa untuk grup eksperimen 79,20, dan hasil proses belajar siswa untuk grup kontrol 76,15. Nilai paling tinggi dari grup eksperimen 96 dan nilai paling rendah diraih 60, namun untuk nilai paling tinggi dari grup kontrol 92 dan nilai paling rendah diraih 56. Analisis data dilaksanakan mengacu terhadap hasil post-test semua sampel yang diawali melalui pengujian normalitas. Uji kenormalan data memakai uji Lilivors untuk mendapatkan hasil yakni sampel didapatkan dari populasi yang sebaran datanya normal. Nilai Lo ditentukan melalui temuan pengujian normalitas yang dilaksanakan. Harga Lo ini selanjutnya dibanding terhadap $L_{tabel}$ pada tingkat signifikansi 0,05. Data yang didapatkan melalui *post-test* kedua grup sampel mempunyai nilai $L_o < L_t$ pada taraf nyata 0,05 yakni pada grup kontrol 0,14 < 0,17 dan grup eksperimen 0,12 < 0,16. Hal tersebut memperlihatkan kedua grup sampel mempunyai sebaran data yang normal. Hasil proses hitung uji kenormalan data lengkapnya bisa diperhatikan pada Tabel 4.

| Kelas | $\alpha$ | N | $L_0$ | $L_t$ | Keterangan |
|---|---|---|---|---|---|
| Eksperimen | 0,05 | 30 | 0,12 | 0,16 | Normal |
| Kontrol | 0,05 | 27 | 0,14 | 0,17 | Normal |

Tabel 4 memperlihatkan dimana pada grup eksperimen yang totalnya 30 siswa dan grup kontrol yang totalnya 27 siswa meraih data distribusinya normal yang didapatkan melalui hasil post-test semua grup sampel. Hal tersebut memperlihatkan dimana kedua grup sampel meraih angka $L_o < L_t$ pada taraf nyata 0,05 yakni grup kontrol 0,14 hingga 0,17 dan kelas eksperimen 0,12 < 0,16.

Data ini dipakai dalam proses uji hipotesis melalui penggunaan uji-t. Sebelum mengadakan pengujian hipotesis, kita laksanakan uji homogenitas. Pengujian ini dilaksanakan dengan tujuan meninjau untuk semua grup sampel didapatkan dari populasi yang jenisnya sama. Standar deviasi untuk grup eksperimen yang total anaknya 30 siswa yakni 10,63. Standar deviasi 27 pada grup kontrol diraih 9,44. Hasil pengujian homogenitas yang dilaksanakan didapatkan 1,91, melalui dk pembilang 29 dan dk penyebut 27, memperlihatkan angka $F_h$ =1,27 dan $F_t$=0,05. Hasil tersebut memperlihatkan dimana Fh <.F(0,05);(30,27). Maknanya untuk semua kelompok data memiliki jenis data yang sama. Hasil proses hitung secara lengkapnya bisa diperhatikan pada Tabel 5.

| Kelas | N | X | S | $S^2$ | Fh | Ft | Dk | Keterangan |
|---|---|---|---|---|---|---|---|---|
| Eksperimen | 30 | 79,20 | 10,63 | 112,99 | 1,27 | 1,91 | 29 | Homogen |
| Kontrol | 27 | 76,15 | 9,44 | 89,21 | | | 26 | |

Berdasarkan Tabel 5 terlihat bahwa Fh < Ft menunjukkan varians yang homogen pada kedua kelas sampel. Sesudah dilaksanakan pengujian kenormalan data dan homogenitas, kedua kelas sampel terbukti berdistribusi yang normal dan mempunyai varian yang sejenis. Oleh karena itu, pengujian kesamaan dua rata-rata dengan menggunakan uji-t digunakan untuk menguji hipotesis. Hasil Perhitungan Data Uji Hipotesis t Independen Hasil post-test dapat dijelaskan dengan fakta dimana pada taraf nyata 0,05, hasil proses uji diraih nilai $t_h$ = 1,14 dan $t_t$ = 2,00. Hasil proses hitung memperlihatkan dimana nilai $t_h$ diraih dalam rentang penyisihan $H_0$. Berdasarkan hasil tersebut maka $H_l$ diterima. Maknanya dijumpai perbedaan pada nilai proses belajar untuk semua sampel, sehingga bisa didapastkan hasil yakni penerapan model generatif dibantu dengan media *mind mapping* bisa memberikan pengaruh pada hasil belajar siswa. Hasil proses hitung pengujian hipotesis lengkapnya ditampilkan pada Tabel 6.

**Tabel 6. Hasil Uji Hipotesis Menggunakan Uji-t**

| Kelas | N | $\alpha$ | X | S | $S^2$ | $t_h$ | $t_t$ |
|---|---|---|---|---|---|---|---|
| Eksperimen | 30 | 0,05 | 79,20 | 10,63 | 112,99 | 1,09 | 2,00 |
| Kontrol | 27 | 0,05 | 76,15 | 9,73 | 94,68 | | |

Tabel 6 memperlihatkan rata-rata nilai pada grup eksperimen yang totalnya 30 siswa diraih 79,20 dan rata-rata nilai untuk grup kontrol yang totalnya 27 siswa yakni 76,30. Standar deviasi pada grup eksperimen didapatkan 112,99 dan standar deviasi pada grup kontrol didapatkan 94,68.

Hasil pengujian pada level aktual 0,05 memperlihatkan nilai $t_h$ = 1,09 dan $t_t$ = 2,00. Hasil proses hitung ditaih nilai $t_h$ berada pada rentang penekanan Ho. Berdasarkan hasil tersebut maka $H_l$ diterima. Maknanya dijumpai perbedaan hasil proses belajar untuk semua grup sampel, sehingga bisa diraih hasil akhir yakni penerapan model generatif yang dibantu dengan media *mind mapping* bisa memberikan pengaruh pada hasil belajar dari siswa.

Penilaian hasil belajar siswa pada aspek keterampilan dilakukan pada saat kegiatan diskusi sedang berlangsung. Terdapat empat indikator keterampilan yang dinilai selama pembelajaran. Empat indikator yang dinilai yang mencakup atas mengamati, menanya, mengolah informasi, dan melakukan komunikasi memakai lembar penilaian diskusi. Nilai rata-rata setiap indikator menyebabkan adanya perbedaan nilai kemampuan dari grup eksperimen dan kontrol. Nilai keterampilan kelas eksperimen menunjukkan lebih unggul dibandiing terhadap grup kontrol.

**Pembahasan**

Dari hasil analisis data dan uji hipotesis mengenai aspek kognitif dan kemampuan diidapatkan dua temuan penelitian. Temuan yang didapatkan dari penelitian ini yakni untuk meninjau tujuan dari penelitian ini bisa diraih. Hasil dari penelitian menampilkan ditemukan perbedaan dari hasil proses belajar yang diraih sesudah memakai model ajar tipe Generatif yang dibantu dengan media *mind mapping* untuk grup eksperimen, dan model yang biasa pendidik gunakan melalui bantuan media *mind mapping* pada kelas kontrol. Berdasarkan hasil analisis memperlihatkan dimana rata-rata nilai *post-test* siswa pada grup eksperimen lebih unggul dibanding akan siswa dari grup kontrol. Peningkatan rata-rata nilai *posttest* pada kedua kelas sampel menunjukkan bahwa siswa mulai lebih termotivasi untuk belajar. Besarnya nilai rata-rata dari grup eksperimen dibanding terhadap grup kontrol, disebabkan pelaksanaan dalam penerapan model ajar tipe Generatif yang dibantu dengan media *mind mapping*.

Model ajar tipe Generatif yang dibantu dengan media *mind mapping* bisa mempermudah siswa dalam belajar sehingga dapat meningkatkan pemahaman siswa. Meningkatnya pemahaman siswa ini karena penerapan model pembelajaran Generatif diawali dengan sintaks *Orientation* yang berisi tentang bagaimana pendidik mempersiapkan peserta siswa dan memberikan motivasi. Siswa disiapkan dengan memeriksa kehadiran kemudian dilanjutkan memberikan motivasi berupa fakta-fakta yang berhubungan dengan materi pembelajaran. Sintaks *Conflict Cognitive* berisi tentang bagaimana pendidik memberikan suatu *Conflict Cognitive* kepada siswa. *Conflict Cognitive* berisi tentang informasi penting yang membantu siswa untuk mengenali miskonsepsi dengan memberikan stimulus. Stimulus yang diberikan yaitu berupa fakta dan permasalahan yang ditemukan dalam kehidupan seharisehari. Selanjutnya pendidik memberikan pertanyaan-pertanyaan yang menantang. Sintaks *Disclosure* yaitu pendidik membantu peserta didik menemukan solusi permasalahan yang ditemukan pada tahap *Conflict Cognitive*. Pada tahap ini, siswa mengungkapkan ide-ide untuk memecahkan masalah berdasarkan pengetahuan awal dan pengetahuan baru. Sintak *Construct* pada tahap ini pendidik mengkonstruksi pengetahuan siswa hal tersebut dilakukan melalui kegiatan eksperimen. Siswa bisa menghubungkan pengetahuan dasar terhadap pengetahuan yang baru didapatkan sehingga membentuk sebuah konsep. Sintaks aplikasi berfokus pada pemecahan masalah. Siswa dapat menggunakan akumulasi pengetahuan dalam pemecahan permasalahan fisika. Pada tahap ini, siswa harus menerapkan apa yang telah dipelajarinya untuk membangun pengetahuan dan memecahkan masalah yang diberikan umpan balik. Sintaks *Reflection Evaluation*, Pendidik menyajikan umpan balik terhadap anak supaya menunjang peningkatan hasil belajarnya [3].

Berdasarkan analisis data hasil belajar pengetahuan dari post-test didapatkan angka rata-rata dari grup eksperimen 79,20, dan nilai rata-rata dari grup kontrol 76,15. Perbedaan pencapaian hasil belajar pengetahuan disebabkan adanya perbedaan dari model ajar yang dipakia untuk tiap grup sampel. Model ajar tipe generatif yang dibantu dengan media *mind mapping* sangat cocok untuk membantu siswa menemukan informasi yang bermakna, mempertanyakan konsep siswa, menarik perhatian, mengidentifikasi miskonsepsi, dan memberikan motivasi pada siswa, akan menjadikan siswa akan lebih mudah untuk paham dengan materi dinamika gerak. Melalui pembelajaran generatif, proses pembelajaran di kelas menjadi lebih aktif, memungkinkan siswa menghubungkan isi pembelajaran dengan peristiwa dan fenomena kehidupan sehari-hari, serta membantu mereka mencapai tujuan belajarnya [9].

Hasil belajar aspek kemampuan pada grup eksperimen lebih unggul dibanding terhadap grup kontrol. Melalui penggunaan model ajar dalam fisika, siswa dapat memperoleh pemahaman lebih dalam terhadap materi pelajaran. Ada empat indikator keterampilan yang dinilai pada saat pembelajaran. Empat indikator yang dinilai adalah observasi, menanya, mengolah informasi, dan keterampilan komunikasi. Hasil pembelajaran akan ditentukan dengan menggunakan lembar evaluasi diskusi. Hasil belajar keterampilan siswa ditentukan oleh rubrik penilaian. Penelitian memperlihatkan dimana penerapan model ajar tipe generatif memberikan dampak. Dampak ini mencakup atas dampak pengajaran dalam bentuk hasil dari proses belajar yang terukur dan dampak yang mengiringi yakni y hasil belajar. Hasil dari uji hipotesis menampilkan penerapan model ajar tipe generatif menyumbangkan pengaruh pada kompetensi pengetahuan dan keterampilan siswa. Hal ini terlihat dari ketuntasan belajar siswa ketika pra dan pasca dilaksanakan pengajaran generatif, Hasil pengujian hipotesis menunjukkan bahwa penerapan model pembelajaran generatif berpengaruh terhadap kompetensi pengetahuan dan keterampilan siswa. Hal ini terlihat dari ketuntasan belajar siswa ketika pra dan pasca dilaksanakan proses belajar secara generatif, serta ketuntasan belajar klasikal siswa setelah dilaksanakan pembelajaran generatif jauh lebih tinggi [8].

**SIMPULAN**

Berdasarkan dari temuan penelitian, penerapan model ajar generatif yang dibantu dengan media media *mind mapping* terbukti memberikan dampak terhadap kompetensi pengetahuan dan keterampilan siswa. Pembelajaran generatif berbantuan media *mind mapping* membantu siswa lebih memahami konsep dan meningkatkan hasil belajar siswa.

**DAFTAR PUSTAKA**